\documentclass[preprint,showpacs,preprintnumbers,amsmath,amssymb]{revtex4}
\usepackage{amsmath,amssymb,graphicx}
\usepackage{epstopdf}
\usepackage {amssymb}
\newcommand{\nc}{\newcommand}
\nc{\ba}{\begin{eqnarray}}
\nc{\ea}{\end{eqnarray}}
\newcommand\be{\begin{equation}}
\newcommand\ee{\end{equation}}

\nc{\x}{{\bf{x}}} \nc{\y}{{\bf{y}}} \nc{\f}{{\bf{f}}}
\nc{\vo}{{\bf{v}}} \nc{\p}{{\bf{p}}}

\input{epsf.sty}

\begin{document}

\title{Cosmic Strings Collision in Cosmological Backgrounds }

\author{Hassan Firouzjahi$^{1}$}
\email{firouz(AT)ipm.ir}
\author{Salomeh Khoeini-Moghaddam$^{1,2}$}
\email{skhoeini(AT)tmu.ac.ir}
\author{Shahram Khosravi$^{2,3}$}
\email{khosravi(AT)ipm.ir}

\affiliation{$^{1}$ School of Physics, Institute for Research in
Fundamental Sciences (IPM), P. O. Box 19395-5531, Tehran, Iran}
\affiliation{$^{2}$ Department of Physics, Faculty of
Science,Tarbiat Mo'allem university, Tehran, Iran }
\affiliation{$^{3}$School of Astronomy, Institute for Research in
Fundamental Sciences (IPM), Tehran, Iran}

\date{\today}
\begin{abstract}

The collisions of cosmic strings loops and the dynamics of junctions formations in expanding backgrounds are  studied. The key parameter controlling the dynamics of junctions formation, the cosmic strings zipping and unzipping is the relative size of the loops compared to the Hubble expansion rate at the time of collision.  We study analytically and numerically these processes for large super-horizon size loops, for small sub-horizon size loops as well as for loops with the radii comparable to the Hubble expansion rate at the time of collision.

\end{abstract}

\maketitle

\section{Introduction}

In models of brane inflation cosmic strings are copiously produced 
\cite{Sarangi:2002yt, Majumdar:2002hy}, for reviews see e.g. 
\cite{Copeland:2009ga, HenryTye:2006uv, Kibble:2004hq, ACD, Mairi}. These cosmic superstrings are in the forms of fundamental strings (F-strings), D1-branes (D-strings) or the bound states of $p$ F-strings and $q$ D-strings, the $(p,q)$ strings. When two $(p,q)$ cosmic superstrings collide junctions are formed due to charge conservation. This is in contrast to the collision of conventional gauge strings where upon collision they exchange partners and intercommute with the probability close to unity. Therefore one may consider the junction formation as a novel feature of 
a network of cosmic superstrings which may prove crucial in cosmic superstrings detection  in cosmological observations. 
Networks of cosmic strings with junctions have interesting physical properties, 
such as the formation of multiple images \cite{Shlaer:2005ry, Brandenberger:2007ae}
and non-trivial gravitational wave emission \cite{Brandenberger:2008ni, Leblond:2009fq}.
Different theoretical aspects of $(p,q)$ string construction were studied in  \cite{Copeland:2003bj, Firouzjahi:2006vp, Jackson:2004zg, Firouzjahi:2007dp, Cui:2007js, Davis:2008kg}
while the cosmological evolution of a string network  with junctions has been investigated in 
\cite{Tye:2005fn}. 

In a recent paper  \cite{Firouzjahi:2009nt}
the collision of two loops of cosmic strings in a flat background was
studied. It was found that with appropriate initial conditions determined by the angle of collision, the colliding loops velocities and the loops relative tensions, junctions can form. However, after the junction is formed it can not grow indefinitely and after some time the junction start to unzip and the colliding loops  disentangle and pass by from each other. The  junctions' zipping and unzipping are interesting and yet non-trivial dynamical properties.  These phenomena becomes more significant in the light of cosmic strings simulation by 
Urrestilla and Vilenkin \cite{Urrestilla:2007yw}. In their model, the cosmic strings  are two types of U(1) gauge strings with interactions between them. 
Due to the interaction, the strings cannot exchange partners and a bound state
will form if the strings are not moving too fast.
It was shown that the length and the distribution of the string network are dominated by the original  
strings and there is a negligible contribution to the string network length and population  from the bound states strings. 
This can be understood based on the following two reasons. Firstly,  the junctions may not form if the colliding strings are moving very fast  so  they can simply pass through 
each other \cite{Bettencourt:1994kc, Copeland:2006eh, Copeland:2006if, Copeland:2007nv, Salmi:2007ah, Bevis:2008hg, Achucarro:2006es}. Secondly and more interestingly, if the junctions are formed, they start to unzip during the evolution.  

Our aim here is to generalize the results of \cite{Firouzjahi:2009nt} to the case of cosmic strings loops collision in cosmological backgrounds, i.e. the radiation and the matter dominated era. As we shall see in sections \ref{junction} and \ref{numeric}, the size of the loops compared to the Hubble radius at the time of collision plays a significant role in junctions evolutions and 
cosmic strings zipping and unzipping. 

The paper is organized as follows. In section 
\ref{formalism} we present our set up and provide the formalism of junction formation 
for arbitrary cosmic strings loops colliding  in cosmological backgrounds. This is a generalization of \cite{Copeland:2006eh} and \cite{Bevis:2009az} where they presented the formalism of cosmic strings collision in the flat background.
In section \ref{junction} we concentrate to the example of two identical loops in cosmological backgrounds. After setting the background equations for the loops profiles, we present the equations governing the dynamics of the junctions. For the case of very large loops (super-horizon size loops) and very small loops (sub-horizon size loops) 
we are able to present some analytical results. In section \ref{numeric} we present our full numerical results for different loop configurations. The conclusion is given in section \ref{conclusion}

\section{Loops collision in  an expanding universe}
\label{formalism}

Here we present the formalism of junction formation 
for arbitrary cosmic strings loops colliding in a cosmological background. In section  \ref{junction}
we employ the results obtained in this section to the particular example of two identical loops at collision in cosmological backgrounds.
The formalism of cosmic strings collision in a flat background was studied in \cite{Copeland:2006eh} and \cite{Bevis:2009az}. 

Our cosmological background is the standard FRW metric
 \ba
 \label{metric}
  ds^2=a^2(\tau)(d\tau^2-d\x^2) \, ,
\ea
where $\tau$ is the conformal time related to the cosmic time $t$ via $d t = a d \tau$,
$a(\tau)$ is the scale factor and we assume that the background space-time has no spatial curvature. Our  cosmological background is 
either radiation dominated (RD)
or matter dominated (MD). 

Suppose
$X^{\mu}_i$ represents the profile of the $i$-th cosmic string in
the target space-time.  As usual, we can go to the temporal gauge
where the time on the string world-sheet is the same as the
conformal time,  $X^0= \tau$, and $X^\mu_i = ( \tau, \x_i)$.
Denoting the other coordinate of the world-sheet parameterization
by $\sigma$, the gauge condition
\ba
 \label{gauge} \dot \x_i \cdot
\x'_i =0 \, , \ea
holds  where the $\cdot$ and the prime indicate the
derivatives with respect to $\tau$ and $\sigma$ respectively.

\begin{figure}[t]
\vspace{-2.8cm}
   \centering
    \includegraphics[width=7in]{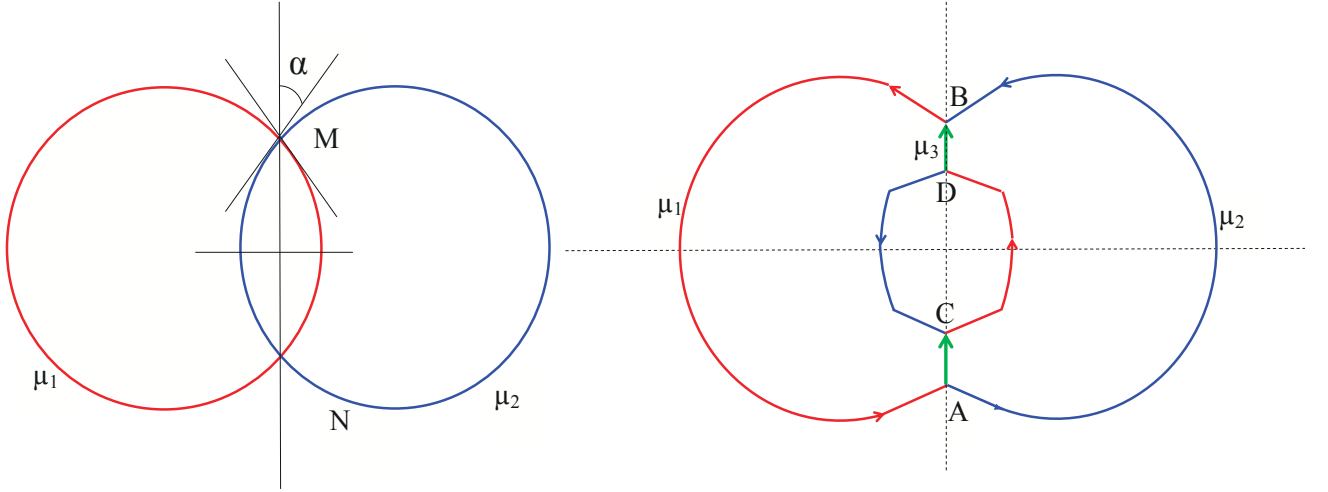}
    \hspace{0cm}
    \vspace{-4cm}
\caption{ A schematic view of the loops at the time of collision(left) and after collision (right). The
arrows in the right figure indicate the directions in which the $\sigma_i$ coordinate increases.
We use the convention that on a loop $\sigma_i$ runs counter clockwise.  There are four junctions
and eight kinks in total.
 }
\vspace{0.8cm}
\label{loop-collision}
\end{figure}

A schematic view of two loops of cosmic strings  in collision is shown in
{\bf Fig. \ref{loop-collision} }.
After collision, there are four junctions and eight kinks. The formation of the kinks is a manifestation of the fact that the speed of light is finite and  parts of the old strings which did not
``feel'' the formation of junctions evolve as before.
In the following we denote the incoming strings by $\x_i$ where $i=1,2$ whereas the newly formed strings are denoted by $\y_a$ where
$a=1,2,3$.  The junctions and the kinks on each string
are described by
$\sigma_a=s_a(\tau)$  and $\sigma_i=\omega_i(\tau)$ respectively.


As mentioned in \cite{Firouzjahi:2009nt, Bevis:2009az}
one complexity of dealing with loops in collision is the orientation of the $\sigma_i$ coordinate
at junctions. We follow the prescription of \cite{Bevis:2009az} and use the sign parameterization
for $\delta_a^J$ according to which $\delta_a^J$ can take values $\pm 1$. If the value of $\sigma_a$
of a particular string increases(decrease) towards the junction $J$, we assign $\delta_a^J=+1(\delta_a^J=-1)$.
With this prescription, the two ends of a piece of string ending in two neighboring junctions have
opposite $\delta$ parameters. The arrows in  {\bf Fig. \ref{loop-collision}} indicate this prescription.
Since it is important for the later analysis, we now give the values of $\delta_a^J$ at each junction:
\ba
\label{delta-value}
A:  \left|
\begin{array}{c}
\delta_1=+1 \\
\delta_2 =-1\\
\delta_3=-1
\end{array}
\right. \hspace{0.5cm}
B:  \left|
\begin{array}{c}
\delta_1=-1 \\
\delta_2 =+1\\
\delta_3=+1
\end{array}
\right.
 \hspace{0.5cm}
C:  \left|
\begin{array}{c}
\delta_1=-1 \\
\delta_2 =+1\\
\delta_3=+1
\end{array}
\right. \hspace{0.5cm}
D:  \left|
\begin{array}{c}
\delta_1=+1 \\
\delta_2 =-1\\
\delta_3=-1
\end{array}
\right.
\ea

Equipped with the $\delta$-prescription, the action for the system of two loops in collision is

\ba\label{kinks-action}
 \mathbf{S}&=&-\sum_J\sum_{i=1}^2\mu_i\int
 d\tau d\sigma_ia^2(\tau)\sqrt{(1-\dot{\y}_i^2) \, {\y'_{i}}^2}
 \theta(\delta_i^J(s_i^J(\tau)-\sigma_i)) \,
 \theta(\delta_i^J(\sigma_i-\omega_i^J(\tau)))\\\nonumber
 &-&\sum_{i=1}^2\mu_i\int
 d\tau d\sigma_ia^2(\tau)\sqrt{(1-\dot{\x}_i^2) \,
 {\x'_{i}}^2}\theta(\delta_i^{A}(\omega_i^{A}(\tau)-\sigma_i)) \,
 \theta(\delta_i^{B}(\omega_i^{B}(\tau)-\sigma_i))\\\nonumber
 &-&\sum_{i=1}^2\mu_i\int
 d\tau d\sigma_ia^2(\tau)\sqrt{(1-\dot{\x}_i^2) \,
 {\x'_{i}}^2}\theta(\delta_i^{C}(\omega_i^{C}(\tau)-\sigma_i)) \,
 \theta(\delta_i^{D}(\omega_i^{D}(\tau)-\sigma_i))\\\nonumber
 &-&\mu_3\int
 d\tau d\sigma_3a^2(\tau)\sqrt{(1-\dot{\y}_3^2) \, {\y'_{3}}^2}
 \theta(\delta_3^A(s_3^A(\tau)-\sigma_3)) \,
 \theta(\delta_3^C(s_3^C(\tau)-\sigma_3))\\\nonumber
 &-&\mu_3\int
 d\tau d\sigma_3a^2(\tau)\sqrt{(1-\dot{\y}_3^2) \, {\y'_{3}}^2}
 \theta(\delta_3^B(s_3^B(\tau)-\sigma_3)) \,
 \theta(\delta_3^D(s_3^D(\tau)-\sigma_3))\\\nonumber
  &+&\sum_J\sum_{a=1}^3\int
 d\tau
 a^2(\tau) \, \f_a^J \cdot[ \, \y_a(s_a^J (\tau),\tau)-\bar{\y}^J(\tau)] \\\nonumber
&+& \sum_J\sum_{i=1}^2\int d\tau a^2(\tau) \, {\bf k}_i^J \cdot[
\, \x_i(\omega_i^J(\tau), \tau)-\y_i(\omega_i^J(\tau), \tau) \, ]
\ea where J  represents the junctions A, B, C and D collectively. Here $\f_a^J$ and ${\bf k}_i^J$ are Lagrange multipliers which enforce that at the kinks the newly formed strings and the old strings meet, $\x_i(\omega_i^J(\tau), \tau) = \y_i(\omega_i^J(\tau), \tau)$ and on the junctions the three newly formed strings join together  $\y_a(s_a^J (\tau),\tau)=\bar{\y}^J(\tau)$ where 
$\bar{\y}^J(\tau)$ represents the position of the junction $J$ in target space.
 As described above, in our convention
 $\delta_i^J$ is +1 if $\sigma$ increases towards the
junction and -1 in the opposite case.

Varying the action  with respect to $\f_a^J,  {\bf
  k}_a^J$ and $\bar \y_a^J$, respectively,  results in
\ba
 \y_a(s^J_a(\tau),\tau)=\bar{\y}^J(\tau)\label{leqm1} \\
 \x_i(\omega_i^J(\tau),\tau)=\y_i(\omega_i^J(\tau),\tau)\label{leqm2} \\
 \sum_a\f_a^J=0\label{leqm3} \, .
\ea

Varying the action with respect to $\x_i$ results in the following standard equations \cite{Turok:1984db}
for the segments of old strings extended between two nearby kinks which are not influenced by the junctions formations
 \ba
  \frac{\partial}{\partial\tau}(\dot{\x}_i\epsilon_{\x_i})+2\frac{\dot{a}}{a}\dot{\x}_i\epsilon_{\x_i}=
  \frac{\partial}{\partial\sigma}(\frac{\x'_i}{\epsilon_{\x_i}}) \, ,
  \label{leqm4}
  \ea
whereas matching the Dirac delta functions gives the following boundary conditions at the kinks $\sigma_i = \omega_i^J(\tau)$
  \ba
  {\bf
  k}_i^J=\mu_i \left(\dot{\x}_i\epsilon_{\x_i}\dot{\omega}_i^J\delta_i^J+\frac{\x'_i}{\epsilon_{\x_i}}\delta_i^J \right) \quad , \quad
  \sigma_i = \omega_i^J(\tau) \, .
  \label{leqm5}
 \ea
 Here 
 $\epsilon_i$ is defined by \cite{Turok:1984db}
  \ba
  \epsilon_i \equiv \sqrt{\frac{\x_i'^2}{1- \dot \x_i^2}}
  \ea
Similarly, varying the action with respect to $\y_a$ results in the following equations for the
newly formed strings stretched between a junction and the nearby kink
  \ba
  \frac{\partial}{\partial\tau}(\dot{\y}_a\epsilon_{\y_a})+2\frac{\dot{a}}{a}\dot{\y}_a\epsilon_{\y_a}=
  \frac{\partial}{\partial\sigma}(\frac{\y'_i}{\epsilon_{\y_i}}) \, .
  \label{leqm6}
  \ea
Now there are Dirac delta functions at the junctions $\sigma_a = s_a^J(\tau)$ and the kinks (for strings 1 and 2) $\sigma_i= \omega_i^J$  which result
   in the following boundary conditions
  \ba
  {\bf k}_i^J=\mu_i \left(\dot{\y}_i\epsilon_{\y_i}\dot{\omega}_i^J\delta_i^J+\frac{\y'_i}{\epsilon_{\y_i}}\delta_i^J \right)
  \quad , \quad \sigma_i= \omega_i^J(\tau)
   \label{leqm7a}         \\
  \label{leqm7}
  \f_a^J=\mu_a \left(\dot{\y}_a\epsilon_{\y_a}\dot{s}_a^J\delta_a^J+\frac{\y'_a}{\epsilon_{\y_a}}\delta_a^J \right)  \quad , \quad \sigma_a = s_a^J(\tau) \, .
  \label{leqm8}
 \ea
Combining Eqs  (\ref{leqm3}) and (\ref{leqm8}) result in
 \ba\label{leqm10}
  \sum_a\mu_a\delta_a^J\left(\dot{\y}_a\epsilon_{\y_a}\dot{s}_a^J+\frac{\y'_a}{\epsilon_{\y_a}}\right)=0 \, .
 \ea

Eliminating ${\bf k}_i^J$ from Eqs. (\ref{leqm5}) and (\ref{leqm7a}), and using the gauge condition (\ref{gauge}), it is easy  to show that
 \ba
  \epsilon_{\x_i}=\epsilon_{\y_i}=\epsilon_i\\
  \delta^J_i\dot{\omega}_i^J\epsilon_i=-1
 \ea

The solutions of the loops in cosmological backgrounds can not be expressed in terms of the the usual right- and left-movers. However, one can define
the right- and left-momenta $\p_a^\pm$ as
 \ba
 \label{def-momenta}
  {\p^\pm}_{\y_a}^J=\frac{\y'_a}{\epsilon_a}\pm \delta_a^J\dot{\y}_a\nonumber \\
  {\p^\pm}_{\x_i}^J=\frac{\x'_i}{\epsilon_i}\pm \delta_i^J\dot{\x}_i
 \ea
 with ${\p_a^\pm}^2=1$.
 Starting with the time derivative of (\ref{leqm1})
\ba\label{leqm2p}
 \x'\dot{\omega}_i^J+\dot{\x}_i=\y'\dot{\omega}_i^J+\dot{\y}_i \, ,
\ea
one  can show that ${\p^-}_{\y_i}^J={\p^-}_{\x_i}^J$. This is the key formula which relates the unknown quantities ${\p^-_{\y_i}}^J$ for the newly formed strings 
to the known quantities ${\p^-_{\x_i}}^J$ from the old strings.

Starting with the time-derivative of Eq. (\ref{leqm1}) combined with Eq. (\ref{leqm10})
one obtains an equation for $\p_{\y_a}^+$ in terms of $\p_{\y_b}^-$
\ba
 \delta_a^J ( 1 + \delta_a ^J \epsilon_a \dot s_a^J ) \,  \p_{\y_a}^+ =
\delta_a^J ( 1 - \delta_a ^J \epsilon_a \dot s_a^J ) \,
\p_{\y_a}^- - \frac{2}{\bar \mu}    \sum_b  \mu_b  \delta_b^J  ( 1
- \delta_b ^J \epsilon_b \dot s_a^J )  \, \p_{\y_b}^- \, .
\ea
Imposing the condition ${\p_a^\pm}^2=1$ one obtains
\ba\label{eqm19}
 q_a- q_a\sum_{b}
 \bar \mu_bq_b \, c_{ab}+\sum_{bc} \bar \mu_b \bar
 \mu_c q_b q_c \, c_{bc} -1=0
\ea
 where $\bar \mu_a \equiv \mu_a/\bar \mu$, $\bar \mu \equiv \sum_a \mu_a$, $q_a^J \equiv 1-\delta_a^J\epsilon_a\dot{s}_a^J$ and
 $c_{ab} \equiv \delta_a^J\delta_b^J\p^-_{\y_a} \cdot \p^-_{\y_b}$.

Now we provide an equation for the energy conversation at the junction.
 For this purpose, multiplying
 Eq (\ref{eqm19}) by $\bar \mu_a$ and summing over $a$ results in
 \ba
\sum_a  \bar \mu_a q_a - \sum_{ab} \bar \mu_a \bar \mu_b \, c_{ab}
+\sum_{abc}  \bar \mu_a \bar \mu_b \bar \mu_c  q_b q_c\, c_{ab} 
-1=0
 \ea
However, with the reshuffling of the indices, one can easily check that the second and the third term above cancel out and one obtains
 \ba\label{energy}
 \sum_a\bar\mu_aq_a=1
\ea
 or
 \ba
 \label{energy}
 \sum_a \delta_a^J \mu_a \epsilon_a \dot s_a =0 \, .
 \ea
 This is a generalization of the case of strings in flat background studied in \cite{Copeland:2006eh} corresponding to
 $\epsilon_a =0$.

 Defining $\hat{c}_{ab}=1-c_{ab}$ and using the energy conservation one can check that
 Eq. (\ref{eqm19})  results in
\ba
 q_a \sum_b  \bar \mu_b \hat c_{ab}  q_b    = \sum_{bc} \bar \mu_b \bar \mu_c \hat c_{bc} q_b q_c
\ea
 Writing explicitly,  and using (\ref{energy}),  yields
\ba
 q_1(1-2\bar \mu_1)(\bar \mu_2q_2 \hat c_{12}+\bar \mu_3q_3 \hat c_{13})= 2 \bar \mu_2 \bar \mu_3q_2q_3 \hat c_{23}\\
 q_2(1-2\bar \mu_2)(\bar \mu_1q_1 \hat c_{12}+\bar \mu_3q_3 \hat c_{23})= 2 \bar \mu_1 \bar \mu_3q_1q_3 \hat c_{13}\\
 q_3(1-2\bar \mu_3)(\bar \mu_2q_2 \hat c_{23}+\bar
 \mu_1q_1 \hat c_{13})= 2 \bar \mu_2 \bar
 \mu_1q_2q_1 \hat c_{12} \, .
\ea
Eliminating $q_3$ and $q_2$ from the first two equations above
and plugging in (\ref{energy}) results in our main formula of interest
\ba
\label{s-eq}
1- \delta_1^J
 \epsilon_1 \dot s_1^J = \frac{ \bar \mu M_1 \hat c_{23}}{\mu_1
 \left[ M_1 \hat c_{23}+ M_2 \hat c_{13}+ M_3 \hat c_{12} \right]}
 \, ,
\ea
where $M_1 \equiv \mu_1^2 - (\mu_2 - \mu_3)^2$ with a similar definition for
$M_{2}$ and $M_3$. One can also obtains a similar equation for $\dot s_{2,3}$ with
an appropriate permutation of the indices. This set of equations for $\dot s_a^J$ is our starting point to study the evolutions of junctions.


\section{Junction evolutions}
\label{junction}
In previous section we have presented the general formalism of cosmic strings loops collision in a cosmological background. Here we specialize to the example of two identical loops at collision in a cosmological background where the analysis can be handled somewhat analytically.

Before dealing with the loops in collision, here we summarize the background solutions for
a loop in expanding background.  Suppose the collision happens at the time $\tau=\tau_0$.  One can check that the
cosmic time $t$ and the conformal time $\tau$ are related by
$t = \tau^{n+1}/\tau_0^n (n+1)$
where we have considered a power law expansion for the scale factor $a(\tau) =
( \tau/\tau_0)^n$. For a radiation an matter dominated universe, $n=1,2$ respectively.
In this convention, the scale factor at the time of collision is equal to unity.
Also, calculation the Hubble expansion rate, $H= d a/a \, dt$, one can check that the Hubble
expansion rate at the time of collision is $H_0 = n/\tau_0$.

Consider a loop extended in $x-y$ plane moving relativistically in $z$ direction. We choose the following ansatz for
the loop configuration
\ba
\label{loop-backg1}
\x= \left(
\begin{array}{c}
 f(\tau) \cos  \frac{\sigma}{R_0}   \\
f(\tau) \sin \frac{\sigma}{R_0}      \\
 z(\tau)
\end{array}
\right ) \, .
\ea
In this picture, $R(\tau) \equiv a (\tau) f(\tau) $ is the physical radius of the loop
and $R_0 = f(\tau_0)$ represents the size of the loop at the time of collision.

The independent equations of motions are
\ba
\label{back-eq1}
F'' &+& \frac{2n}{x} F' ( 1- v^2 - F'^2) + ( 1- v^2 - F'^2) F^{-1} =0 \\
\label{back-eq2} v' &+& \frac{2n}{x} v ( 1- v^2 - F'^2) =0 \, .
 \ea
Here the loop center of mass velocity is defined by $v = \dot z(\tau)$. For the ease
of the numerical investigations, we introduced the dimensionless
time variable $x \equiv \tau/\tau_0$ and $F(x) \equiv
f(\tau)/\tau_0$. Also the prime here and below represents
derivatives with respect to the dimensionless time $x$.

This definition leads
to $F= f H_0/n$ which has a simple  physical interpretation as follows. In our convention
$a(x) = x^n$, so at the time of collision, corresponding to $x=1$, $a=1$. The physical
radius of the loop at the time of collision therefore is $R_0= f(x=1)$. Therefore, the initial condition $F(x=1) = f(x=1) H_0/n$ is a measure of the physical radius of the loop compared to the Hubble radius at the time of collision. For Loops of super-horizon size at the time of collision $F(x=1) >1$, whereas for small sub-horizon sized loops at the time of collision
$F(x=1) <1$. In the following, to simplify the notation we set $F(x=1) \equiv F_1$.

In general it is not easy to find analytical solutions for the set of equations (\ref{back-eq1})
and (\ref{back-eq2}). We solve this equation numerically. Our goal is to calculate $\hat c_{ab}$ for the loop with ansatz (\ref{loop-backg1}) and then obtain the evolution of junctions $\dot s_a^J$. However, one may get some useful analytical
information in some certain limits.  These include very large super-horizon size loops at
the time of collision ($F_1 \gg 1$) and small sub-horizon size loops ($F_1 \ll 1$).

The collisions of loops in Minkowski background was studied in  \cite{Firouzjahi:2009nt}.
Here we generalize that study to the case of strings loops collision in an expanding background.
In order to simplify the analysis, we consider the symmetric case where the two incoming loops have equal tension and physical radius at the time of collision.  We assume that the loops are extended in $x-y$ plane and are moving along the $z$-direction with velocity $\pm v$. A schematic view of this example is given in {\bf Fig.} \ref{loop-collision}  .
By symmetry the newly formed string 3 will be static extended either along $x$ or $y$ directions. Whether it is a $x$-link or a $y$-link junction depends on the angle of collision
$\alpha$ \cite{Copeland:2006eh}. For small enough angle of collision it is a $y$-link while for a large enough angle of collision it would be an x-link junction.
To be be specific, we consider a $y$-link junction where the string 3 is extended along the $y$-direction.

The profiles of the colliding loops are given by
\ba\label{ex-eq1}
 \x_i=\left (\begin{array}{c}
\mp b+ f(\tau) \cos  \frac{\sigma_i}{R_0}   \\
 f(\tau) \sin \frac{\sigma_i}{R_0} \\
 \pm z(\tau)
 \end{array}\right)
\ea
 where for i=1 (i=2) we  choose upper sign (lower sign). Here $2b$ is the impact factor.

From the continuity of the left moving momenta one has ${\p_{\y_i}^{-}}^J={\p_{\x_i}^{-}}^J$
which can be served to find ${\p_{\y_i}^{-}}^J$. From (\ref{def-momenta}) we have
 \ba\label{ex-eq2}
 {\p_{\y_i}^{-}}^J={\p_{\x_i}^{-}}^J=\left (\begin{array}{c}
 -\sqrt{1-F'^2-v^2}\sin \frac{\sigma_i}{R_0}-\delta_i^J F'\cos \frac{\sigma_i}{R_0}   \\
\sqrt{1-F'^2-v^2}\cos \frac{\sigma_i}{R_0}-\delta_i^J F'\sin \frac{\sigma_i}{R_0}  \\
 \mp \delta_i^J v(\tau)
 \end{array}\right)
\ea

Using $\hat c_{ab} ^J\equiv 1 - \delta_a^J \delta_b^J  {\p_{\y_a}^-}^J \cdot {\p_{\y_b}^-}^J$
at each junction and the fact that $\delta_1^J=-\delta_2^J=-\delta_3^J$, one obtains
\ba
 \label{c_{ab}}
 \hat c_{11}^J&=&\hat c_{22}^J=\hat
 c_{33}^J=0\\
 \hat c_{12}^J&=&1+ v^2 - (1-2F'^2-v^2)\cos 2 S_1^J+ 2\delta_1^J F' \sqrt{1-F'^2-v^2}\sin 2 S_1^J \label{c12}\\
 \hat c_{13}^J&=&1+\sqrt{1-F'^2-v^2}\cos S_1^J-\delta_1^J F'\sin S_1^J \label{c13}\\
 \hat c_{23}^J&=&1-\sqrt{1-F'^2-v^2}\cos S_2^J-\delta_1^J F' \sin S_2^J\label{c23}\\
 \label{c123}
 \hat c_{23}^J+\hat c_{13}^J&=&2+2\sqrt{1- F'^2-v^2}\cos S_1^J-2\delta_1^J F'\sin S_1^J
\ea
Here and below, to simplify the notation the definition $S_a^J \equiv s_a^J/R_0$ is introduced.
We note that $S_a^J$ is dimensionless which is more suitable for numerical analysis.
To obtain Eqs. (\ref{c12}-\ref{c23}) we note that due to symmetry in problem, one has
$ S_1^{'J} = - S_2^{'J}$ at each junction $J$.   On the other hand, one also observes that
$S_1^B(x=1) +S_2^B(x=1) =  \pi $ and $S_1^A(x=1) +S_2^A(x=1) =  3\pi  $
which was used to simplify the final results in Eqs. (\ref{c12}-\ref{c23}).

Plugging  $\hat c_{ab}$ in our master equation (\ref{s-eq}), the evolution of the junction is given by
\ba
\label{s3J-eq}
{S_3'}^J = \frac{\delta_3^J}{F_1}\frac{( \kappa - 1) (\hat c_{13} + \hat c_{23} ) - \kappa\,  \hat c_{12} }
{  ( \kappa - 1) (\hat c_{13} + \hat c_{23} ) +   \hat c_{12}  } \, ,
\ea
where the dimensionless parameter $\kappa$ is given by the ratio of the tensions 
$\kappa\equiv 2 \mu_1/\mu_3 $.

Also from the energy conservation Eq. (\ref{energy}) one has
\ba
\label{energy3}
 {S_1'}^{ J}= \frac{ F_1}{\kappa F}  \sqrt{1-F'^2-v^2} \,   {S_3'}^{ J} \,  .
\ea

Equations (\ref{s3J-eq}) and (\ref{energy3}) jointly can be used to solve for 
${S_3'}^{ J} $. Due to symmetries involved in the problem, we only need to find the evolution of junctions $B$ and $D$ and the evolutions of junctions $A$ and $C$ are mirror images
of junctions $B$ and $D$. We note that the conditions for the junction formation  is that the 
string 3 stretching between  junctions $B$ and $D$ to be created. This requires that its length to increase initially with time: $S_3' >0$ where $S_3 \equiv S_3^B - S_3 ^D$ measures the length of string $BD$. As we shall see in our numerical results, usually this is translated into 
${S_3^B}'>0$ and ${S_3^D}'<0$. However, in some very fine-tuned situations one can also 
find examples where ${S_3^B}'>0$ and ${S_3^D}'>0$ such that $S_3' >0$ is still satisfied.

As explained in \cite{Firouzjahi:2009nt}, after the junction formation, the entangled loops start to unzip. The onset of unzipping at junction $J$ happens when $ {S_3'}^{J}$
vanishes and changes its sign. As we shall see later, 
the unzipping times for junctions $B$ and $D$
are not equal. Since we are mainly interested in the evolution of the newly formed string $BD$, we define the onset of unzipping for string $BD$ when $S_3$ reaches a maximum and $S_3'=0$. After that the length of string $BD$ reduces with time.    
Sometime after unzipping, the loops disentangle from each other and pass by in opposite directions. The time of loops disentanglement happens when the junctions B and D meet
corresponding to $S_3=0$. However, we also encounter examples where the loops shrink to zero before they disentangle from each other.

As explained above, the onset of unzipping at junction $J$ is determined when ${ S_3'}^J=0$.
Here we show that the denominator in Eq. (\ref{s3J-eq}) 
is always positive so the sign of ${ S_3'}^J$ is controlled by the numerator of the above expressions.
To see that note that
$  -\sqrt{p^2 + q^2}  \leq  p \, cos \theta + q \sin \theta \leq \sqrt{p^2 + q^2}$ for real numbers
$p$ and $q$ and arbitrary angle $\theta$. Using these inequalities, one can easily check that $\hat c_{12} \ge 2 v^2 $ and  $\hat c_{13} + \hat c_{23} \ge 2 ( 1- \sqrt{1- v^2})$. On the other hand, as described in \cite{Copeland:2006eh}, one also requires that $2\mu_1 > \mu_3$ for the junction formation to be allowed kinematically. In conclusion the denominator in
${ S_3'}^J$ expression is always positive and the sign of ${ S_3'}^J$ evolution is determined by the numerator of the above expressions. This plays important rules in determining the junctions
unzipping times in the following discussions.

As explained before, our goal is to solve the background loop equations (\ref{back-eq1})
and (\ref{back-eq2}) and use the resulting values of $F(x)$ and $v(x)$ 
in $\hat c_{ab}$ expressions to find the junctions evolutions from Eq. (\ref{s3J-eq}).
This procedure can be done only numerically because both $F(x)$ and $v(x) $ can not 
be found analytically in general.
Before presenting our full numerical analysis, we consider two different limits where some analytical insights can be obtained for the junctions evolutions.

\subsection{Large super-horizon size loops}

Here we consider the limit where the colliding loops are much larger than the Hubble radius
at the time of collision, $F_1 \gg 1$.
For the super-horizon size loops, one expect that they are conformally stretched as the universe expands, $R(\tau) \propto a(\tau)$, where $F'$ and $F''$ are small compared to unity. We will demonstrate this in our numerical analysis.  Due to damping effects from the expanding background, the super-horizon loops become non-relativistic and $v \ll 1$.
In this approximation, one can easily solve the background equations (\ref{back-eq1})
and (\ref{back-eq2}) . Denoting $F= F_1 + \Delta$ where $\Delta$ represents the small
evolution of $F$, one obtains
$$\Delta '' + \frac{2}{x} \Delta' + \frac{1}{F_1} \simeq 0 \, ,$$
which after neglecting the sub-dominant term, results in
\ba\label{Delta-eq}
 \Delta \simeq -\frac{x^2}{2(1+ 2 n) F_1}   \quad
, \quad F' \simeq -\frac{x}{(1+ 2 n) F_1} \, .
\ea

As the universe expands, the loop reenter the horizon in its subsequent evolutions. This can be approximated when
$\Delta \simeq - F_1 $ which results in the time of the loop horizon reentry $x_*$
\ba
\label{x-star}
x_* \simeq  \sqrt{2 ( 1+ 2 n)} F_1 \, .
\ea
Our full numerical analysis, as we shall see in next section, verify that this is indeed a good approximation. From this expression for $x_*$ we see that with similar initial conditions, it takes longer for the loops to reenter the horizon  in a matter dominated era
as compared to the radiation dominated era.

As the super-horizon size loops stretches conformally, its center of mass velocity reduces rapidly. For time smaller than $x_*$ one can find an approximate solution for $v(x)$. Neglecting the terms containing $v^2$ and $F'^2$ in Eq. (\ref{back-eq2}) one obtains $v' + 2 n\, v /x \simeq0 $ which easily can be solved to give
\ba
\label{v-evolution}
v(x) \simeq \frac{v_1}{x^{2n}} \,  ,
\ea
where $v_1$ is the value of $v(x)$ at the time of collision, $v_1 \equiv v(x=1)$. As explained above, we see that the loop central mass velocity reduces rapidly with time. We also see that for the matter dominated background with $n=2$, the loss of velocity is more pronounced as compared to the radiation dominated background with $n=1$.

For the super-horizon size loops and keeping only terms up to $F'$ in $\hat c_{ab}$
and neglecting $F'^2$ and $v^2$ as explained above, one obtains
\ba
\hat c_{12} \simeq 1- \cos  2 S_1^J  + 2 \delta_1^J F'
\sin  2 S_1^J \quad , \quad
\hat c_{13}+ \hat c_{23} \simeq 2+ 2 \cos S_1^J
 - 2  \delta_1^J F'   \sin S_1^J  \nonumber
\ea
Plugging these into ${S_3'}^J$  expression results in

\ba \label{s3J2-eq} {S_3'}^J \simeq \frac{\delta_3^J}{F_1} \frac{
 \left( 1+  \cos   S_1^J \right) \left( \kappa \cos   S_1^J
-1   \right) -  \delta_1^J F' \sin  S_1^J
\left( \kappa - 1 +  2 \kappa \cos   S_1^J   \right) } {  
 \left( 1+ \cos S_1^J  \right)  \left( - \cos  S_1^J +\kappa  \right )   -  \delta_1^J F' \sin  S_1^J \left( \kappa - 1 - 2  \cos   S_1^J    \right) } \, .
 \ea
At the time of collision, one can neglect the terms containing $F'$ in above expression and for small $x$ one obtains
\ba
S_3^{'B}({x \simeq 1}) \simeq \frac{\delta_3^J}{F_1}  \frac{
  \kappa  \cos  S_1^J   -1    }
{ \kappa -     \cos   S_1^J    } \, .
\ea
For the junction to form we need ${S_3^B}'(x=1)>0$.
With $S_1^{B}(x=1)=\alpha$, the condition for junction formation is  translated into 
\ba
\label{alphac}
0 \leq \alpha  \leq \alpha_c \quad , \quad \alpha_c = \arccos (\kappa^{-1}) \, .
\ea
Interestingly, this is identical to the bound obtained in \cite{Copeland:2006if} for  straight strings in collision  at the small velocity approximation. This is expected, since the super-horizon size loops can be locally well approximated by the straight strings. Our full numerical analysis, presented in next section, indeed show that the bound on $\alpha_c$ given by
the above equation works very accurately for the large loops.

As time goes by, the terms containing $F'$ in Eq. (\ref{s3J2-eq}) becomes important.
We note that the denominator in  Eq. (\ref{s3J2-eq}) is positive so the sign of $S_3^J$ evolution is determined by the  numerator of Eq. (\ref{s3J2-eq}). Consider junction B for example.
For the junction B to unzip, $S_3^B$ should slow down, requiring that the term containing $F'$ in numerator of Eq. (\ref{s3J2-eq}) gives a negative contribution.
With $\delta_1^B =-1$ and $ F'<0$ the $F'$ correction in numerator of Eq. (\ref{s3J2-eq}) indeed contributes negatively. This indicates that as time goes by, the rate of evolution of $S_3^B$ slows down until ${S_3'}^B=0$ when the junction B starts to unzip.
Similar argument applies to junction D too.

\subsection{Small sub-horizon size loops}
\label{sub-loops}

Now we consider the limit where the colliding loops are much smaller than the Hubble radius
at the time of collision, $F_1 \ll 1$. In this limit, the damping terms in
Eqs. (\ref{back-eq1}) can be neglected \cite{Turok:1984db} and the loop evolution is the same as in the flat background. In this limit $v$ is nearly constant and the loop has a simple periodic profile 
\ba
F(x) \simeq F_1 \cos \left(  \frac{x-1}{\gamma F_1}  \right) \, ,
\ea
where $\gamma = 1/\sqrt{1- v^2}$ is the Lorentz factor. To simplify the analysis, here we chose the initial configuration  such that $F'=0$ at the time of collision. Neglecting the effects of expansion, one would expect the criteria for zipping, unzipping and the loops disentanglement would be similar to cosmic strings loops collision in  a flat background studied in \cite{Firouzjahi:2009nt}.

Starting from the energy conservation formula (\ref{energy3}), one obtains
\ba
S_1^{B(D)} = \frac{ S_3^{B(D)}  }{\kappa \gamma} + \alpha \, .
\ea

Also calculating $\hat c_{ab}$ yields
\ba
\hat c_{12} &=& 2 - 2 \gamma^{-2} \cos^2 \left(   S_1^J + \delta_3^J \frac{x-1}{\gamma F_1}
\right) \nonumber\\
\hat c_{12} + \hat c_{13} &=& 2 + 2 \gamma^{-1} \cos \left(   S_1^J + \delta_3^J \frac{x-1}{\gamma F_1}  \right) \, .
\ea
Plugging these in ${S_3'}^J$ expression yields
\ba
{S_3'}^J = \frac{\delta_3^J}{F_1}  \, \frac{ \kappa \cos \left(   \frac{ S_3^B}{\kappa \gamma }  +  \frac{\delta_3^J(x-1)}{\gamma F_1} + \alpha  \right)  -\gamma  }
{ \kappa  \gamma  - \cos \left(   \frac{ S_3^B}{\kappa \gamma}  +  \frac{\delta_3^J(x-1)}{\gamma F_1} + \alpha  \right)    } \, .
\ea

The details of the loops zipping, unzipping and disentanglement
were studied in \cite{Firouzjahi:2009nt}. Here we briefly outline
the main results.  For the junctions to form one requires that $0
< \alpha < \alpha_c$ where $\alpha_c = \arccos ( \gamma/\kappa)$. The unzipping times for junctions B and D, $x_D^u$ and
$x_B^u$, satisfy

\ba
\label{xD-xB}
 x_D^u - x_B^u = 2 \gamma \alpha F_1 \left( 1- \kappa^{-2} \right)^{-1} \left[   1- \frac{1}{\kappa \gamma}
\frac{\sin \alpha}{\alpha} \right] \, . 
\ea
Since $   \sin \alpha/\alpha $ and $ 1/\kappa \gamma$ are always
less than unity, one concludes that $x_D^u > x_B^u$, indicating
that the junction B  which holds the external large arcs unzip
sooner than the junction D which holds the internal smaller arcs.
Although we have proved this only for small sub-horizon loops but our numerical analysis show that this conclusion holds true in general. 

The loops disentangle at the time $x_f$ when $S_3=0$, which is  
given by the parametric relation
\ba
\kappa \gamma  \cos^{-1} \Gamma -
\cos\left(\frac{x_f-1}{\gamma F_1}\right) \sqrt{1-\Gamma^2} -
\kappa \gamma \,  \mu_1 \alpha   + \sin \alpha \,
 =0 \, , \ea
 where
$$
 \Gamma \equiv \left[  \frac{ (x_f-1)   }{\kappa F_1 \sin(\frac{x_f-1}{\gamma
 F_1})}
  \right] \, .
$$

This is an implicit equation for $x_{f}$ which should be solved in
terms of $\kappa, \gamma, \alpha$ and $F_1$. For this to make
sense, we demand that $x_{f} -1 <  \gamma F_1\pi /2 $ before the loops
shrink to zero.

\section{Numerical analysis}
\label{numeric}
In this section we present our full numerical results for different loops configurations. 
To be specific, we consider three examples of (a): large super-horizon size loops with $F_1=100$, (b): intermediate size loops with $F_1 = 0.5$ and (c): small sub-horizon size loops with $F_1=0.01$.

\begin{figure}[t]
   \centering
    \includegraphics[width=3in]{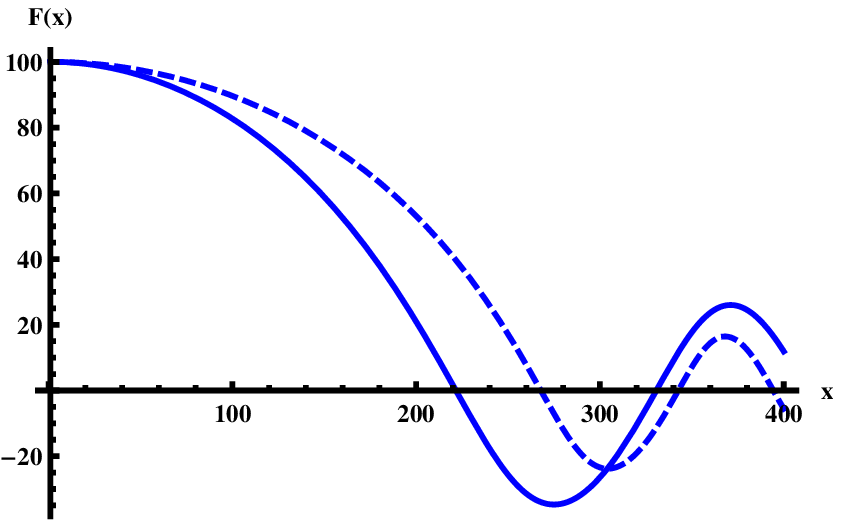} \hspace{0.6cm}
    \includegraphics[width=2.8in]{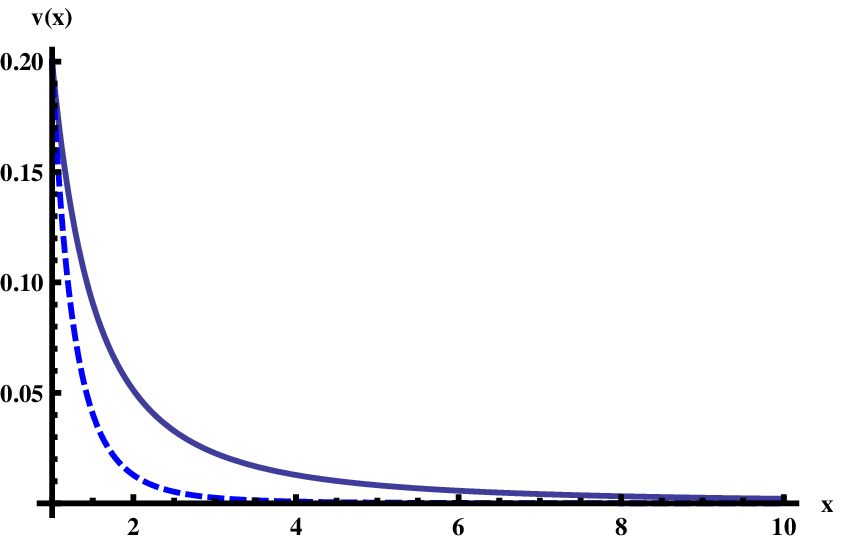}     \vspace{0.1cm}
    \vspace{0cm}
\caption{ 
Here the background evolution of Eqs. (\ref{back-eq1}) and (\ref{back-eq2})
are presented with $ F_1 =100, v_1 =0.2$ and $  F'(x=1)=0.1$. The left figure shows $F(x)$ whereas the right figure represents $v(x)$.
 The solid  lines are for the radiation dominated backgrounds $(n=1)$ and the dashed lines are for the matter dominated backgrounds $(n=2)$.
  We are considering the loops evolution until they shrink, i.e. until the first root of $F(x)=0$.
 }
\vspace{0.5cm}
\label{F1=100-background}
\end{figure}

\begin{figure}[h]
   \centering
    \includegraphics[width=3in]{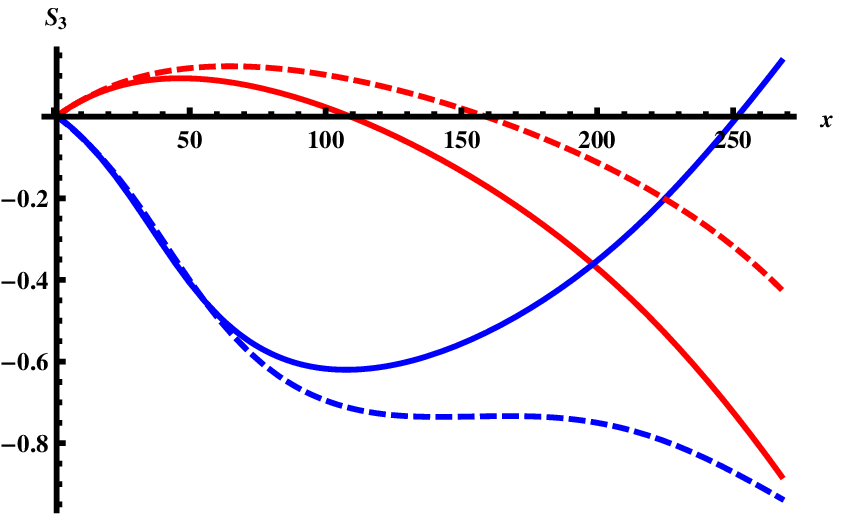} \hspace{0.6cm}
    \includegraphics[width=2.8in]{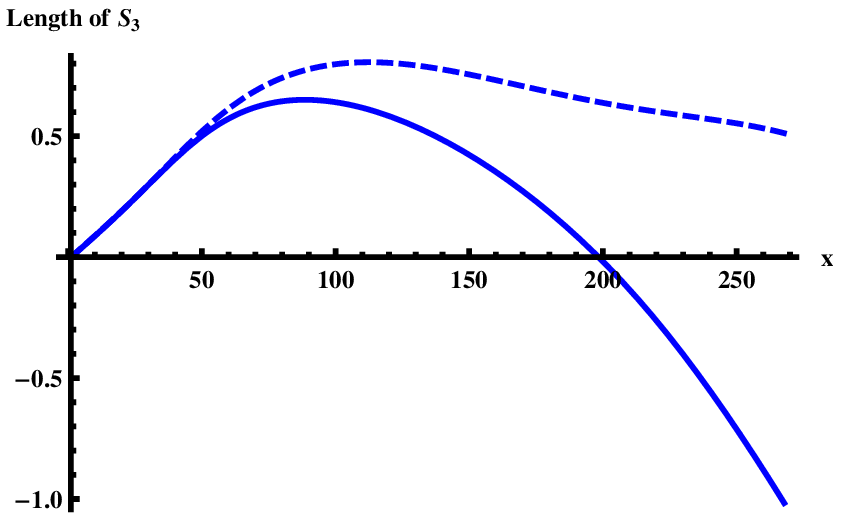}     \vspace{0.1cm}
    \vspace{0cm}
\caption{In these plots we have presented the evolution of
junctions $B$ and $D$ for $F_1=100, \alpha =\pi/9$ and
$\kappa=1.2$. In the left figure the  upper solid red curve
represents $S_3^B$ whereas the lower solid blue curve is that of
$S_3^D$ for the radiation dominated background. The  dashed curve
represents the corresponding curves in the matter dominated
background.  The right graph represents the length of the newly
formed strings $\mu_3$, $S_3 \equiv S_3^B - S_3^D$. The solid
(dashed) curve is for the radiation (matter) dominated background.
We see that in both backgrounds, after junction formation, the
string $BD$ reaches a maximum length and get unzipped. However,
only in the radiation dominated background $S_3$ becomes zero
before loops shrink  indicating the loops disentanglement.}
\label{F1=100-junction}
\vspace{1cm}
\end{figure}

\subsection{Large super-horizon size loops}

For super-horizon size loops with $F_1 \gg 1$ one expects that the loops are conformally stretched until they re-enter the horizon. In this period, the loops lose much of its 
center of mass velocity as demonstrated by Eq. (\ref{v-evolution}). As mentioned before, the loss of velocity is more significant for the matter dominated backgrounds. Also from Eq. (\ref{x-star}) we see that it takes longer for the loop to shrink for the matter dominated backgrounds as compared to the radiation dominated backgrounds. Both of these analytical conclusions were verified in our full numerical investigations. In {\bf Fig.\ref{F1=100-background}} we have presented the background solutions of $F(x)$ and $v(x)$ solving Eqs. (\ref{back-eq1}) and (\ref{back-eq2}) numerically. As is clear from the figure, in a matter dominated background, it takes longer for the loop to shrink. This in turn plays some roles in the junctions evolutions and loops disentanglement. In {\bf Fig.\ref{F1=100-junction}} we have presented the evolutions of junction $B$ and $D$ solving Eq. (\ref{s3J-eq}) numerically. The left figure shows the evolution of junctions $B$ and $D$ both for matter and radiation dominated backgrounds. The right figure shows the length of the newly formed string $\mu_3$ stretching between junctions $B$ and $D$ which is $S_3 \equiv S_3^B - S_3^D$. As mentioned previously  for the junction to form
we require that $S_3'>0$. Form the right figure we see that  the junction is created
in both cosmological backgrounds. After the junction formation, $S_3$ reaches a maximum value indicating the unzipping of the newly formed strings $BD$. After this time, $S_3$ reduces. When $S_3=0$ the junctions $B$ and $D$ meet again and the loops disentangle and pass by from each other.  From the right figure we see that for the radiation dominated background the loops disentanglement indeed take place. However, for the matter dominated background, we see that before loops find the opportunity to disentangle, they shrink to zero.
As explained below Eq. (\ref{xD-xB}) the junction $B$ unzips sooner than junction $D$ which is also demonstrated in the left figure of {\bf Fig. \ref{F1=100-junction}}.

\begin{figure}[t]
   \centering
    \includegraphics[width=3in]{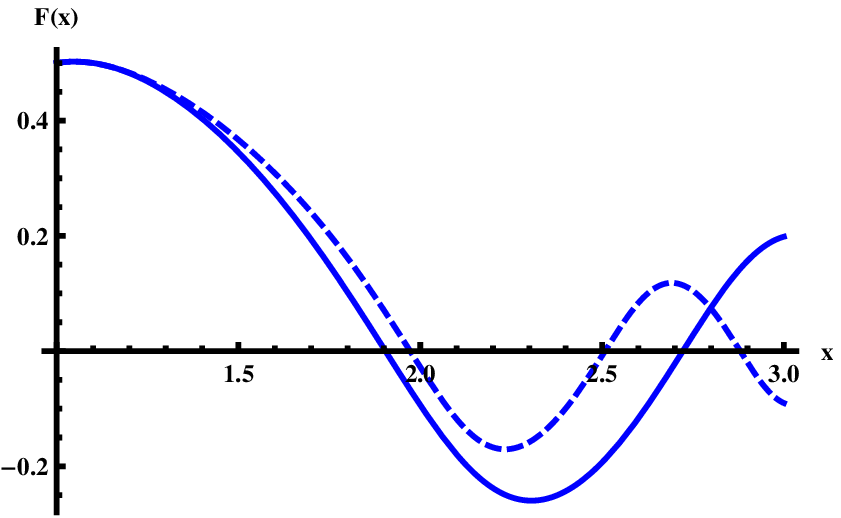} \hspace{0.6cm}
    \includegraphics[width=2.8in]{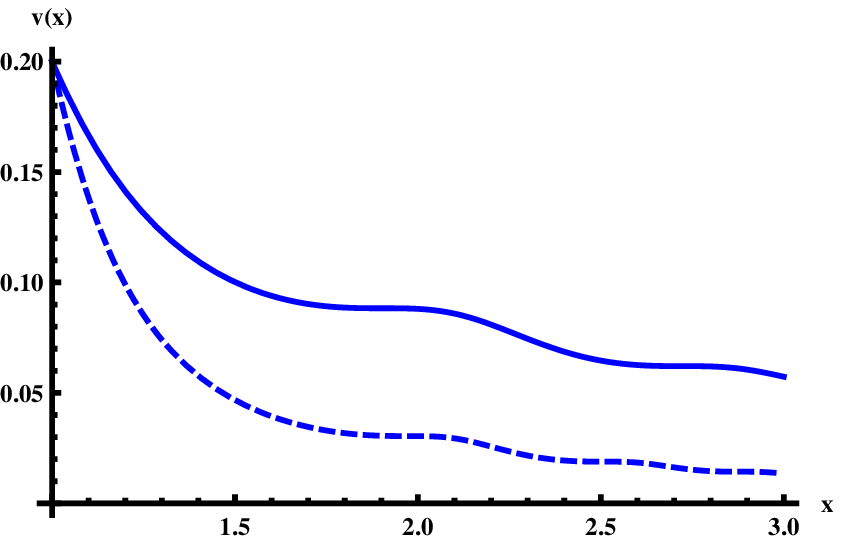}     \vspace{0.1cm}
    \vspace{0cm}
\caption{ Here the background evolution for $F(x) $ and $v(x) $ are presented with $ F_1 =0.5, v_1 =0.2,  F'(x=1)=0.1$.
 The solid  (dashed) curves are for the radiation (matter) dominated backgrounds. }
\vspace{0.8cm}
\label{F1=0.5-background}
\end{figure}
\begin{figure}[h!]
\vspace{0cm}
   \centering
    \includegraphics[width=3in]{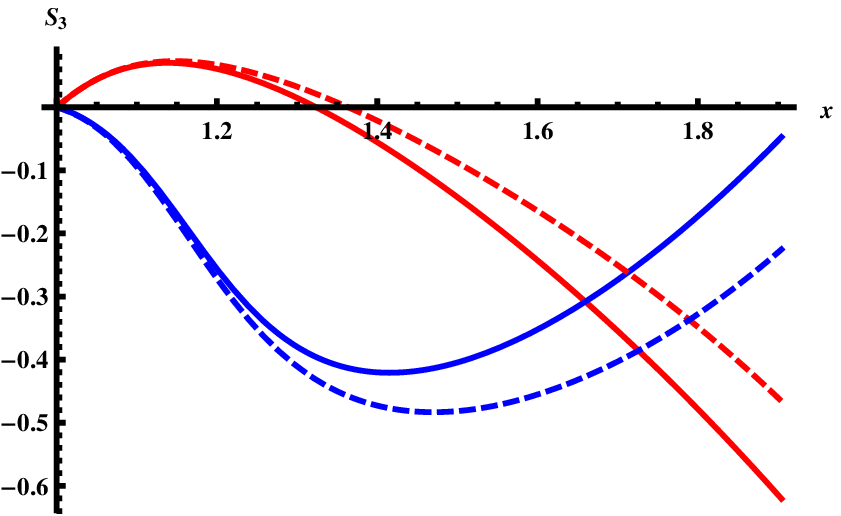} \hspace{0.6cm}
    \includegraphics[width=2.8in]{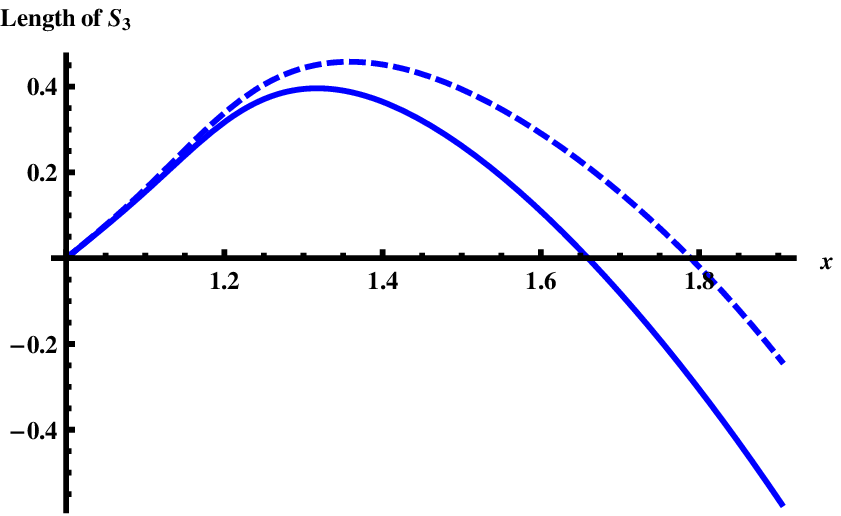}     \vspace{0.1cm}
    \vspace{0cm}
\caption{ Here we present the evolutions of junctions $B$ and $D$
and the string $BD$ for $F_1 =0.5, \alpha =\pi/9$ and  $\kappa
=1.2$. In the left figure the  upper solid red curve represents
$S_3^B$ whereas the lower solid blue curve is that of $S_3^D$
for the radiation dominated background. The  dashed curves
represent the corresponding curves in the matter dominated
background.
 The right graph represents the
length of the newly formed string $BD$, $S_3$. The solid (dashed) curve is for the radiation (matter) dominated background.
 We see that in both backgrounds, after junction formation, the string $BD$ reaches a maximum length and get unzipped.
  We also observe that the onsets of string $BD$ unzipping and the loops disentanglement happen sooner in a radiation dominated era.
 }
\vspace{1cm}
\label{F1=0.5-s3}
\end{figure}

\subsection{Loops with intermediate sizes}
For the loops with the sizes comparable to the Hubble radius at the time of collision, $F_1 \sim 1$, we can only do numerical analysis.
On the physical grounds one expects that the evolution of $F(x)$ and $v(x)$ is less sensitive 
to the background cosmological expansion as compared to large super-horizon size loops.
 In {\bf Fig. \ref{F1=0.5-background}} we have presented the background evolution of $F(x)$ and $v(x)$ for $F_1 = 0.5$. As expected $v(x)$ changes slowly and $F(x)$ evolves similarly for both matter and radiation dominated backgrounds. In  
{\bf Fig. \ref{F1=0.5-s3}} we have presented the evolutions of junctions $B$, $D$ and the string $BD$ . For both radiation dominated and matter dominated backgrounds we see 
that the junctions are formed followed by the string $BD$ unzipping and the loops disentanglement. One observes that the onsets of string $BD$ unzipping (when $S_3$ reaches a maximum) and also the loops disentanglement happen earlier for the radiation dominated era as compared to the matter dominated era. This may be interpreted by noting that for the radiation dominated backgrounds the loops reenter the horizon and shrink sooner  as can be seen qualitatively from Eq. (\ref{x-star}).

\subsection{Small sub-horizon size loops}

\begin{figure}[t]
   \centering
    \includegraphics[width=3in]{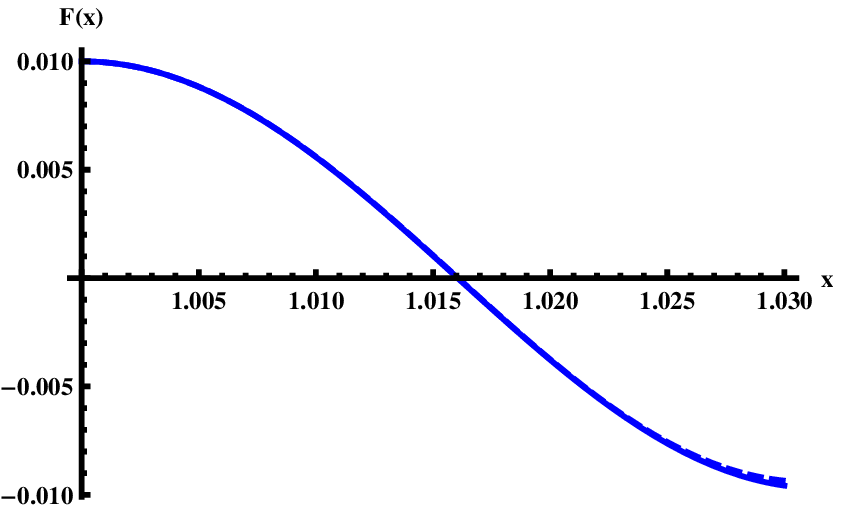} \hspace{0.6cm}
    \includegraphics[width=2.8in]{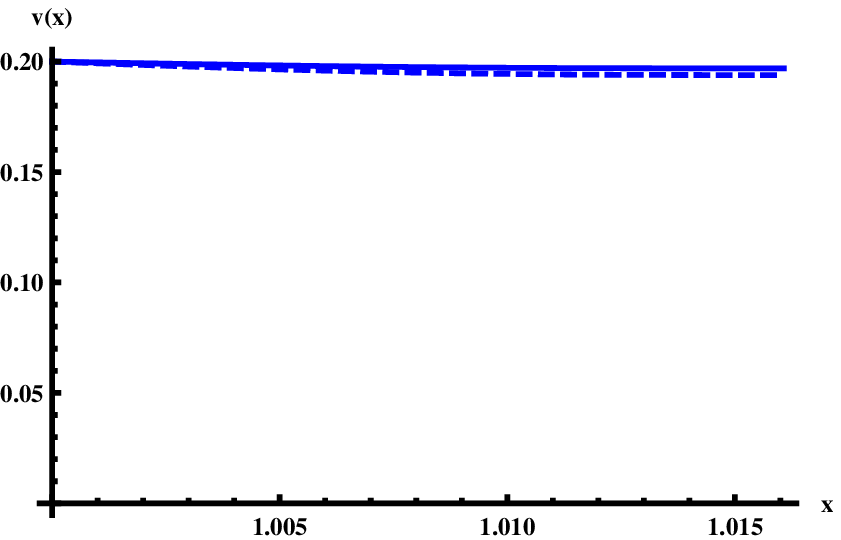}     \vspace{0.1cm}
    \vspace{0cm}
\caption{ Here the solutions of $F(x)$ and $v(x)$ for both matter dominated and radiation dominated backgrounds 
are shown for $F_1 =0.01, v_1 =0.2$ and $ F'(x=1)=0$. As expected, the background  cosmological evolutions do not play important roles so $F(x)$ indicates simple periodic behavior and  $v(x)$ changes slowly.
 }
\label{F1=0.05-background}
\vspace{1cm}
\end{figure}

\begin{figure}[t]
   \centering
    \includegraphics[width=3in]{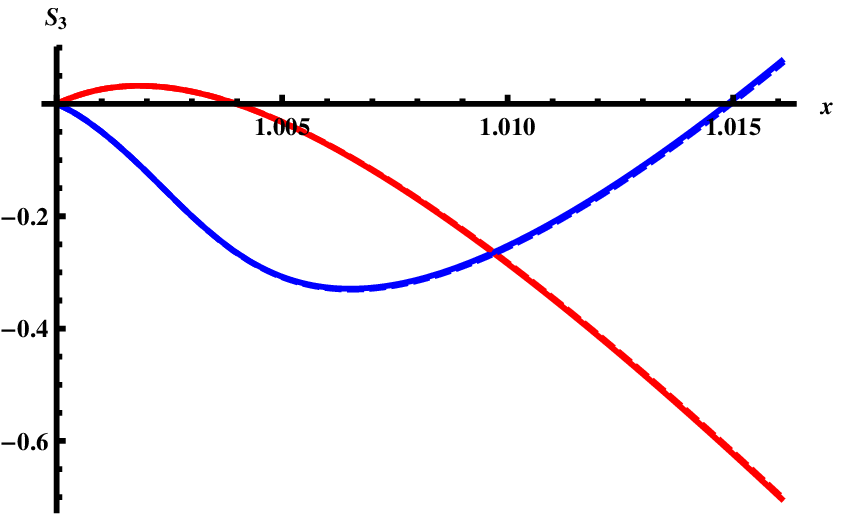} \hspace{0.6cm}
    \includegraphics[width=2.8in]{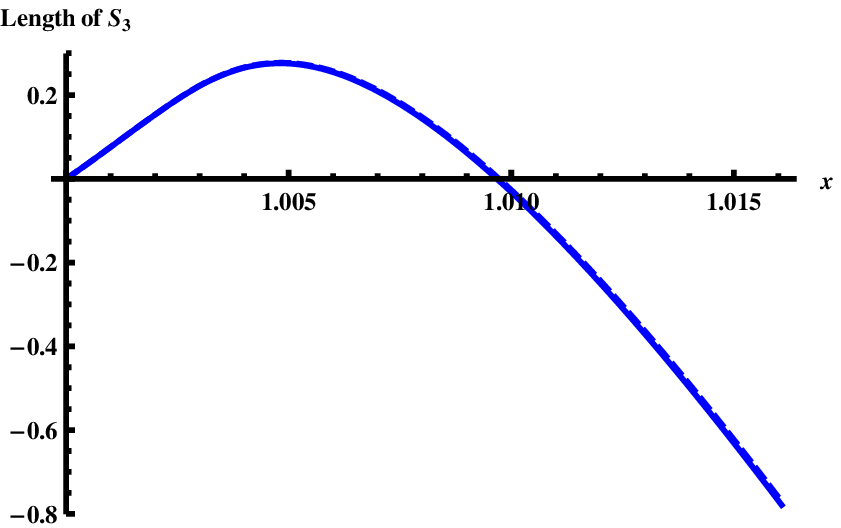}     \vspace{0.1cm}
    \vspace{0cm}
\caption{ Here the junctions evolutions are shown for $ F_1 =0.01,
\alpha=\pi/9$ and $\kappa =1.2$. In the left figure, the upper red
(lower blue) curve shows $S_3^B (S_3^D)$. Interestingly the curves
corresponding to the radiation and the matter dominated
backgrounds coincide to each other. As demonstrated by Eq.
(\ref{xD-xB}) the junction $B$ unzip sooner than the junction $D$.
The right figure shows the length of string $BD$ given by $S_3$.
Again the curves corresponding to the
matter dominated and the radiation dominated backgrounds
coincide.
 }
\vspace{1cm}
\label{F1=0.05-s3}
\end{figure}

For small loops, $F_1 \ll 1$, as explained before one expects the background cosmological evolutions do not play important roles. We have presented the analytical results for small loops
in subsection \ref{sub-loops}. In {\bf Fig. \ref{F1=0.05-background}} we have presented the full numerical solutions of $F(x)$ and $v(x)$. As expected $F(x)$ shows simple periodic behavior and 
$v(x)$ does not change much in each period. In {\bf Fig. \ref{F1=0.05-s3}} we have presented the junctions evolution. As expected, the junctions evolutions are identical for bath matter and radiation dominated backgrounds.  We also observe that the string BD unzipping and the loops disentanglement take place in this example. As demonstrated analytically in Eq. (\ref{xD-xB}) the junction $B$ unzips sooner than junction $D$ which is also demonstrated in the left figure of {\bf Fig. \ref{F1=0.05-s3}}.

\section{conclusion}
\label{conclusion}

In this work we have studied the cosmic strings collision in cosmological backgrounds. After presenting the general formalism in section \ref{formalism} we have concentrated to the example of colliding loops. The motivation for this work was to understand analytically the findings of  simulation in \cite{Urrestilla:2007yw} where it was found that there were little contributions from the bound states strings in their multiple strings network. One can understand this phenomena as follows. For the junctions to develop upon strings collision, some appropriate initial conditions should be satisfied. 
These depends on the relative tensions of the colliding strings, the angle of collision and their relative velocities. Yet the more interesting observation is that even when junctions are created, they can not grow indefinitely and the bound state strings start to unzip. 

As described in \cite{Firouzjahi:2009nt}, for straight cosmic strings at collision the junctions do not unzip once they are materialized. However, for colliding loops in a flat background
the zipping and unzipping generically happen \cite{Firouzjahi:2009nt}.   
The natural question is how sensitive are these results to the expansion of the Universe. 
Here we find some interesting results indicating that the background expansion plays important roles in strings zipping, unzipping and their eventual disentanglement . The key parameter here is the relative size of the loops compared to the Hubble expansion rate at the time of collision. For large super-horizon size loops one may approximate them with straight strings. 
This implies that 
if junctions are formed upon loops collision it will grow initially as in straight strings examples. However, as the Universe expands the loops stretch conformally until they re-enter the Horizon. Meanwhile their velocities reduces rapidly. The net effect is that the rate of bound state strings creation slows down until it starts to unzip. Eventually the loops disentangle from each other and pass by from each other in opposite directions if they did not shrink to zero by then. On the other hand, for small sub-horizon size loops one can neglect the effects of the expansion and the results of \cite{Firouzjahi:2009nt}
holds true. The case of colliding loops with the sizes comparable to the Hubble radius at the time of collision is more non-trivial which shares some features with small and large loops cases.

Our numerical investigations also show that the junction formation and the zipping and unzipping phenomena are sensitive to the angle of collision $\alpha$ and the strings relative tensions parametrized by $\kappa$. It would be interesting to study these phenomena in the multi parameter space of $F_1, \kappa $ and $\alpha$. Also to simplify the analysis, we have restricted ourselves to the example of coplanar 
colliding loops with equal tensions and radii. 
It would be interesting to study the general case where the loops have different sizes and orientations and  there are hierarchies between the sizes of the loops and the Hubble expansion rate at the time of collision.


\section*{ Acknowledgments}

We would like to thank Tom Kibble for useful discussions and comments. 

\end{document}